\documentclass[12pt]{article}

\usepackage[pdftex]{graphicx}

\usepackage{epstopdf,SIunits}
\DeclareGraphicsExtensions{.pdf,.mps,.png,.jpg,.gif}

\begin{document}
\begin{flushright} \today
\end{flushright}

~

\begin{center} \bf \Large
Search for dark matter at high-power laser facilities : 
flawed luminosity calculations in QPS -- Quasi parallel scattering.
\end{center} 

~

\begin{center} 
Denis Bernard

~

{\it
Laboratoire Leprince-Ringuet, Ecole Polytechnique, CNRS/IN2P3
\\
91128 Palaiseau, France
}
\end{center} 

~

Thanks to the tremendous progress made  by high-efficiency high-repetition-rate high-power short-pulse lasers
\cite{Mourou:2013} 
and anticipating the
availability of facilities such as  ELI  \cite{ELI}, a series
of works have been published recently that propose to use lasers to search for
new physics effects such as a  dark matter scalar field $S$, 
in the stimulated collision of laser pulses
\cite{Fujii:2010is,Homma:2011pg,Homma:2012mn,Homma:2012zz}.
The proposed scheme is based on two principles; 
\begin{enumerate}
\item
 The stimulation of a $\gamma \gamma \rightarrow S \rightarrow \gamma
 \gamma $ elastic scattering reaction by the presence in the initial state
 of a pulse of photons that are identical, in the Heisenberg sense, to
 one of the photon expected to be produced in the final state, as was
 proposed in Refs.  \cite{Moulin:2002ya} and
 \cite{Lundstrom:2005za,Lundin:2006wu} and tested experimentally in
 Ref. \cite{Bernard:2010dx};
\item
 The angular configuration proposed
 \cite{Fujii:2010is,Homma:2011pg,Homma:2012mn}, named
 ``Quasi parallel scattering'' (QPS), is the collision of two
 copropagating lasers.  The natural angular divergence of the laser
 provides a range of possible small angles $\theta$ for the collision
 of two particular photons and therefore a range of possible
 center-of-mass system energies for that collision, the range
 overlapping, hopefully, the narrow line resonance of the interaction
 through a scalar $S$ having a given mass \cite{Fujii:2010is}.

The availability of very small angles in this QPS configuration
enables the search for objects with a very small mass
 ($m\ll$ 1\,eV/$c^2$) despite the high photon energy ($E \approx$ 1 eV)
of the laser.
\end{enumerate}

Alas, the expression of the luminosity used, 
eq. (5.1) of Ref.  \cite{Fujii:2010is},
eq. (34) of Ref.  \cite{Homma:2011pg},
eq. (32) of Ref.  \cite{Homma:2012mn},
is the expression for the collision of head-on beams, a configuration
which is far from being the configuration used here.  
The luminosity for the head-on collision of two bunches is :
\begin{equation}
{\cal L}_0 = \frac{C N_1 N_2}{A},
\end{equation}
where $N_1$ and $N_2$ are the number of particles in beam $1$ and
$2$, $A$ is the transverse geometrical area over which the crossing
takes place, and $C$ is a numerical factor of order unity, that
depends of the details of the transverse shapes of the beams.
This expression is often also used for ``small angle crossing'', which in
the context of accelerator physics means an almost head-on collision.

But here, where the two beams are co-propagating, the full expression
\cite{Moller:1945} must be used,
\begin{equation}
{\cal L} = 
 \frac{K}{2}  {\cal L}_0 = 
\frac{K}{2} \frac{C N_1 N_2}{A},
\end{equation}
where $K$ is a ``Kinematic factor'' :
\begin{equation}
K = \sqrt{(\vec v_1 - \vec v_2)^2 - (\vec v_1 \times \vec v_2)^2/ c^2}. 
\end{equation}
Modern discussions of M{\o}ller's opus \cite{Moller:1945} can be found in
Refs.  \cite{Herr:2003em,Furman:2003,Napoly:1992kn}.
Using the QPS $\theta$ angle definition, we get : 
\begin{equation}
K = \sqrt{(2 \sin\theta)^2 - (\sin2\theta)^2},  
\end{equation}
that is  $K/2 =  \sin^2\theta$.
\begin{itemize}
\item For head on collisions, $\theta = \pi/2$ and $K/2=1$, as was
  expected;
\item For $\theta \approx 0$, as is the case here, 
$K/2 = [m/(2 E)]^2$. 
\end{itemize}
The missing  factor, $K/2$, is extremely small for the resonance
masses of interest ($m \ll$ 1\,eV/$c^2$, see Fig.4 of Ref. 
\cite{Homma:2012mn}). 
The potential of  high fluence lasers for exploring  weakly
interacting ``vacuum fields'' such as Dark Matter fields
\cite{Tajima:2013} may have been grossly overestimated.


\begin{thebibliography}{999} \small

\bibitem{Mourou:2013} 
G. Mourou,
``High power Lasers - new developments'',
in
``High-Power Laser Workshop'',  
SLAC 
October 1-2, 2013.

\bibitem{ELI}
The Extreme Light Infrastructure (ELI) 
http://www.extreme-light-infrastructure.eu/

\bibitem{Fujii:2010is} 
  Y.~Fujii and K.~Homma,
  ``An approach toward the laboratory search for the scalar field as a candidate of Dark Energy,''
  Prog.\ Theor.\ Phys.\  {\bf 126}, 531 (2011)
  [arXiv:1006.1762 [gr-qc]].

\bibitem{Homma:2011pg} 
  K.~Homma, D.~Habs and T.~Tajima,
  ``Probing the semi-macroscopic vacuum by higher-harmonic generation under focused intense laser fields,''
 Appl. Phys. B 106:229-240 (2012), 
  arXiv:1103.1748 [hep-ph].

\bibitem{Homma:2012mn} 
  K.~Homma,
  ``Sensitivity to Dark Energy candidates by searching for four-wave mixing of high-intensity lasers in the vacuum,''
 Prog. Theor. Exp. Phys.  04D004 (2012),
  [arXiv:1211.2027 [hep-ph]].

\bibitem{Homma:2012zz} 
  K.~Homma, D.~Habs, G.~Mourou, H.~Ruhl and T.~Tajima,
  ``Opportunities of fundamental physics with high-intensity laser fields,''
  Prog.\ Theor.\ Phys.\ Suppl.\  {\bf 193}, 224 (2012).

\bibitem{Moulin:2002ya} 
  F.~Moulin and D.~Bernard,
  ``Four-wave interaction in gas and vacuum. Definition of a third order nonlinear effective susceptibility in vacuum: $\chi_{vacuum}^{(3)}$,''
  Opt.\ Commun.\  {\bf 164}, 137 (1999)
  [physics/0203069 [physics.optics]].

\bibitem{Bernard:2010dx} 
  D.~Bernard {\it et al.},
  ``Search for Stimulated Photon-Photon Scattering in Vacuum,''
  Eur.\ Phys.\ J.\ D {\bf 10}, 141 (2000)
  [arXiv:1007.0104 [physics.optics]].

\bibitem{Lundstrom:2005za} 
  E.~Lundstrom {\it et al.},
  ``Using high-power lasers for detection of elastic photon-photon scattering,''
  Phys.\ Rev.\ Lett.\  {\bf 96}, 083602 (2006)
  [hep-ph/0510076].

\bibitem{Lundin:2006wu} 
  J.~Lundin {\it et al.},
  ``Detection of elastic photon-photon scattering through four-wave mixing using high power lasers,''
  Phys.\ Rev.\ A {\bf 74}, 043821 (2006)
  [hep-ph/0606136].

\bibitem{Moller:1945}
C. M{\o}ller,
``General Properties of the Characteristic Matrix in the Theory of Elementary Particles I'', 
Danish Academy of Sciences,
Matematisk-Fysiske Meddelelser, Vol. 23, No. 1 (1945) 

\bibitem{Herr:2003em} 
  W.~Herr and B.~Muratori,
  ``Concept of luminosity,''
in 
  ``Intermediate accelerator physics. Proceedings, CERN Accelerator School, Zeuthen, Germany, September 15-26, 2003,''
%

\bibitem{Furman:2003}
 M.~A. Furman, 
``The M{\o}ller Luminosity Factor'',
LBNL-53553, CBP Note-543 (2003).
%

\bibitem{Napoly:1992kn} 
  ``The Luminosity for beam distributions with error and wake field effects in linear colliders,''
  O.~Napoly,
  Part.\ Accel.\  {\bf 40}, 181 (1993).
%

\bibitem{Tajima:2013} 
T. Tajima,
``Overview of High Field Science'',
in
``High-Power Laser Workshop'',  
SLAC 
October 1-2, 2013.
\end{thebibliography}
\end{document}